\documentclass[twocolumn, preprint letterpaper,english,prl,showpacs,superscriptaddress,amsmath,amssymb]{revtex4}
\usepackage[T1]{fontenc}
\usepackage[latin1]{inputenc}
\usepackage{amstext}
\usepackage{graphicx}
\usepackage{color}

\makeatletter
\@ifundefined{textcolor}{}
{%
 \definecolor{BLACK}{gray}{0}
 \definecolor{WHITE}{gray}{1}
 \definecolor{RED}{rgb}{1,0,0}
 \definecolor{GREEN}{rgb}{0,1,0}
 \definecolor{BLUE}{rgb}{0,0,1}
 \definecolor{CYAN}{cmyk}{1,0,0,0}
 \definecolor{MAGENTA}{cmyk}{0,1,0,0}
 \definecolor{YELLOW}{cmyk}{0,0,1,0}
 }
\usepackage{graphicx}
\usepackage{dcolumn}
\usepackage{bm}
\usepackage{amssymb}
\usepackage{amsmath}
\usepackage[matrix,arrow,curve]{xy}
\usepackage{ulem}
\usepackage{psfrag}
\usepackage{subfigure}

\makeatother
\usepackage{babel}
\begin{document}
\title{Optimal waveform for the entrainment of a weakly forced oscillator}

\author{Takahiro Harada}

\affiliation{Department of Physics, Graduate School of Science, The University of Tokyo, Tokyo 113-0033, Japan}

\author{Hisa-Aki Tanaka}

\affiliation{Department of Electronic Engineering, The University of
Electro-Communications, Tokyo 182-8585, Japan}

\author{Michael J. Hankins}

\affiliation{Department of Chemistry, Saint Louis University, St.
Louis, MO 63103.}

\author{Istv\'{a}n Z. Kiss}

\begin{abstract}
A theory for obtaining waveform for the effective entrainment of a
weakly forced oscillator is presented. Phase model analysis is
combined with calculus of variation to derive a waveform with which
entrainment of an oscillator is achieved with minimum power forcing
signal. Optimal waveforms are calculated from the phase response 
curve  and a solution to a balancing condition. The theory is tested in
chemical entrainment experiments in which oscillations close to and
further away from a Hopf bifurcation exhibited sinusoidal and higher
harmonic nontrivial optimal waveforms, respectively.
\end{abstract}

\maketitle
Entrainment of oscillators to an external signal in nonlinear dissipative
systems is a fundamental concept of importance in a large variety
of applications \cite{sync-book}; two prominent examples include
the time-scale adjustment of circadian system to light \cite{winfree-book}
and the cardiac system to a pacemaker \cite{Glass:1991p6559}. The
entrainment process is rigorously described theoretically by phase/amplitude
equations and circle maps for weakly and strongly perturbed nonlinear
systems, respectively \cite{sync-book}. The general result of the
theoretical analysis is that nonlinear oscillators can adjust their
frequencies to that of the external source above a critical forcing
amplitude. In the forcing amplitude vs. forcing frequency diagram
there are long vertical entrainment regions called Arnold tongues.

A widely accepted tool for studying entrainment is the {phase response
curve}  \cite{winfree-book}. {Specifically, the 
phase response function (infinitesimal phase response curve)
} indicates the phase shift of an
oscillator due to an
infinitesimal perturbation of a system variable \cite{Bard}. A classical problem
in nonlinear dynamics uses phase response function and forcing waveform
with which all important features of the entrainment process (e.g.,
locking range, defined as the width of the Arnold tongue at a given forcing amplitude) can be obtained
for weakly perturbed {systems \cite{sync-book}.}

Many applications require optimization of the entrainment process.
This is often achieved by adjusting the forcing waveform to achieve
a target entrainment feature.
A variety of control targets were explored: optimal input was determined
for establishing fast entrainments \cite{Granada:2009p6520}, circadian
phase resetting \cite{Bagheri:2008p6519,Forger:2004p3757}, starting/stopping
of the oscillations \cite{Forger:2004p3757,Lebiedz:2005p6529}, and
maximal resonance (energy transfer)
between the system and forcing signal \cite{Gintautas:2008p6521}. Control 
of deterministic \cite{ASME} and stochastic \cite{Feng-PRL, Ritt-PRE} neuronal spiking activity  was achieved with phase modeling 
approach combined with variational methods to optimize spiking time \cite{ASME} and variance of firing rates \cite{Feng-PRL}.

In this Letter, we give a full account for the inverse of the classical
entrainment problem: what is the minimal power forcing waveform that produces efficient entrainment
of a limit-cycle oscillator in weak forcing limitation? Although the quality of entrainment could involve features such as stability and basin of attraction, here we consider 
efficient entrainment as the occurrence of maximum 
width (or minimum slopes) of the
Arnold tongue.  This {`locking range'} quality marker has been commonly used for
injection-lock micro integrated oscillators as well as phase-locked
loop circuits \cite{Lockrange}. 
We propose a versatile, efficient approach
to obtain exact functional form of the optimal waveform provided that
the response function related to the forcing action had been established.
The theoretically obtained optimal waveforms, which exhibit some unexpected
symmetry relationships with the response function, are tested in a
simple numerical model that include higher harmonics in the 
response function typically seen in strongly nonlinear oscillators. 
The experimental
applicability of the method is demonstrated with optimal entrainment
of a chemical system, the oscillatory electrodissolution of nickel
in sulfuric acid.

The entrainment process of a limit-cycle oscillator
in weak forcing limit can be modeled by \cite{ref-winfree-model}
 \begin{eqnarray}
\frac{d\psi}{dt} = \omega + Z(\psi)f(\Omega t),
\label{eq1}
\end{eqnarray}
where $\psi$ is the phase of the oscillator, $Z$ is the phase response function, and $\omega$ and $\Omega$ are the natural frequency of the
oscillator and the forcing frequency, respectively.
In weak forcing limit, Eq. (\ref{eq1}) is further simplified by averaging \cite{kura}, as
 \begin{eqnarray}
\frac{d\phi}{dt} = \Delta\omega + \Gamma(\phi),
\label{eqnew2}
\end{eqnarray}
where $\phi$ and $\Delta\omega$ are given by $\phi = \psi - \Omega t$, and $\Delta\omega = \Omega - \omega$, respectively \cite{Reso}.
The interaction function, $\Gamma(\phi)$, is  
obtained from the forcing waveform $f$ and the 
{phase response function}, $Z$, as  
{$\Gamma(\phi) = \left \langle
Z(\theta+\phi)f(\theta)\right\rangle$,} 
where $\theta$ represents $\Omega t$ and
$\langle \cdot \rangle$ denotes the average by $\theta$ over its period $2\pi$: {$\left \langle \cdot \right\rangle \equiv (2\pi)^{-1} \oint {\cdot} \hspace{1mm} d\theta$}.  
Entrainment occurs when the phase difference is locked, i.e., 
$d\phi/dt =\Delta \omega + \Gamma (\phi) = 0$ \cite{Forger:2004p3757}.  The range of frequency difference, $\Delta \omega$, where solution for stable steady state exists for $\phi$ defines the locking range ${\it R[f]}$ for a certain
forcing waveform \cite{Stev}. Therefore, the locking range is the difference between
the maximum ({at} $\phi = \phi_{\rm +}$) and minimum ({at} $\phi=\phi_{\rm -}$) {} values 
of $\Gamma(\phi)$ where phase {locked} 
solution exists \cite{R}.      
$R[f]$ is thus given by $\Gamma (\phi_{\rm +}) - \Gamma
(\phi_{\rm -})$. 

Now we are in position to formulate the optimal entrainment 
problem mathematically: the optimal forcing waveform ($f_{{\rm opt}}$) 
maximizes the locking range $R$ under certain constraints. 
A convenient practical constraint is the total power of the waveform 
over its period: $\langle f(\theta)^{2} \rangle$. 
Therefore, the optimal forcing waveform, $f_{{\rm opt}}$ gives 
maximal locking range  for a given (constant) forcing power $P$.  
We consider this as a  variational problem 
maximizing the functional form  
\begin{equation}
\mathcal S[f] \equiv R[f] - \lambda ( \langle f^{2}\rangle-P ),
\label{eq2}
\end{equation}
where $\lambda$ is the Lagrange multiplier.
Solution to the variational problem,  $f_{*}$, a suitable candidate for the optimal 
waveform, is obtained by ensuring that the first variation $\delta \mathcal{S} $  vanishes and the 
second variation $\delta^{2} \mathcal{S} $ is negative \cite{cons}:
\begin{equation}
f_{*}(\theta) = (2\lambda)^{-1} \{ Z(\theta + \phi_{+}) - Z(\theta +
\phi_{-})\}.
\label{eq3}
\end{equation}
The Lagrange multiplier can be obtained by substituting the solution Eq. (\ref{eq3}) in the 
constant power constraint ($\langle f_{*}^2 \rangle - P = 0$): $\lambda = (1/2) \sqrt{Q/P}$ with $Q \equiv \langle \{Z(\theta + \phi_{+}) - Z(\theta + \phi_{-})\}^{2} \rangle$. 
Note that $f_{*}$ in Eq. (\ref{eq3}) has zero average: 
$\langle f_{*} \rangle = 0$.

To obtain  $f_{*}$ of Eq. (\ref{eq3}) the 
maximum ($\phi_{+}$) and the minimum ($\phi_{-}$)  {$\phi$ values} of 
$\Gamma$ with the forcing waveform
\begin{equation}
\Gamma(\phi) = \sqrt{P/Q}\hspace{1mm}\langle Z(\theta + \phi) \{ Z(\theta
+ \phi_{+}) - Z(\theta + \phi_{-}) \} \rangle
\label{eq6}
\end{equation}
have to be determined. The conditions for the maximum and minimum of $\Gamma$ are as follows:
\begin{equation}
\Gamma'(\phi_{\pm}) = 0,\hspace{1mm}
\Gamma''(\phi_{+}) < 0,\hspace{1mm}
{\rm and}\hspace{1mm}\Gamma''(\phi_{-}) > 0.
\label{eq5}
\end{equation}
The first condition in Eqs. (\ref{eq5}), combined with Eq. (\ref{eq6}), gives
\begin{equation}
\langle Z'(\theta + \phi_{+}) Z(\theta + \phi_{-}) \rangle = \langle
Z'(\theta + \Delta \phi) Z(\theta) \rangle = 0
\label{eq7}
\end{equation}
where $\Delta \phi \equiv \phi_{+} - \phi_{-}$. ($\Delta \phi$ is 
introduced to remove phase ambiguity). We shall refer to Eq. (\ref{eq7}) 
as balancing condition because this
equation realizes optimality by balancing both terms in Eq. (\ref{eq3}).
The trivial  $\Delta \phi = 0$ solution to Eq. (\ref{eq7}) is discarded
because it does not allow entrainment ($\Gamma (\phi) \equiv 0$  in Eq. (\ref{eq6})).

However, other solutions do exist in Eq. (\ref{eq7}), because 
$\left. \partial \langle Z'(\theta +
\Delta\phi) Z(\theta) \rangle / \partial\Delta\phi \right|_{\Delta \phi = 0} = -\langle
Z'(\theta)^{2}\rangle<0,$
and $\langle Z'(\theta + \Delta\phi) Z(\theta) \rangle$ is a periodic, 
bounded function of $\Delta \phi$ in a 
large class of systems \cite{sync-book}. 
In particular, we have found that in models with 
twice differentiable, continuous $Z$ a solution
with  $\Delta \phi = \pi$ exists; we call this solution and 
the corresponding optimal waveform {`generic'} \cite{generic}. 

Here we test the above theoretical predictions by a simple model with
the following response function,
\begin{equation}
Z(\theta) = \sin \theta + a \sin( 2 \theta).
\label{eq10}
\end{equation}
This $Z$ simulates the behavior of Stuart-Landau 
oscillator \cite{sync-book} with $a=0$; therefore, we can consider $a$ as a measure of distance 
from Hopf bifurcation that can introduce higher harmonics in the
response function. The balancing condition of Eq. (\ref{eq7}) for this model is explicitly written as
$\left[ 1 + 4 a^2 \cos(\Delta \phi) \right] \sin(\Delta\phi) = 0.$
For $|a| < 1/2$, there is only one (nontrivial) solution $\Delta \phi =
\pi$. 
The optimal waveform for this generic solution is independent of $a$: 
$f_{\rm opt}(\theta) = - \sqrt{2 P} \sin \theta.$
 Although the response function does contain second order harmonic for 
$0<|a|<1/2 $, 
this term does not affect the shape of the optimal waveform. 
This finding suggests that with $Z$ containing only weak second (or, in 
general even) harmonics the sinusoidal forcing is
the optimal since the generic solution to balancing condition always exists. For example, 
for systems close to Hopf bifurcation, which contain mostly first and 
weak second (even) harmonics in $Z$, the optimal waveform is sinusoidal.   However, the odd harmonics do appear in the generic optimal waveform 
and thus systems with relatively strong third (and higher odd) harmonics in $Z$ are expected to retain the odd harmonics in the optimal waveform. 
For the generic solution  the locking range is calculated as 
$R = \sqrt{2 P}$,
which is again independent of $a$. 

In the range $|a| \ge 1/2$, we have three different
solutions for the balancing condition and thus three candidates for optimal waveform.
The generic solution with $ \Delta \phi = \pi$ 
exists, however, there appear two additional, {`non-generic'} solutions
satisfying $\cos \Delta \phi = -1/4a^2$.
For these non-generic solutions the locking ranges are identical:
$R = (1 + 4a^2)\sqrt{P}/(2\sqrt{2}a) \ge \sqrt{2 P}$.
This shows that in the range $|a| \ge 1/2$, the non-generic optimal
waveforms outperform the generic waveform;
for large values of $a$, the improvement 
of locking range  for the non-generic over generic waveforms 
increases approximately linearly.
The non-generic waveforms are not
purely sinusoidal and depend on the parameter $a$.
Two non-generic waveforms for $a=0.95$ are shown in Fig. 1(a). 

\begin{figure}
\includegraphics[scale=0.35]{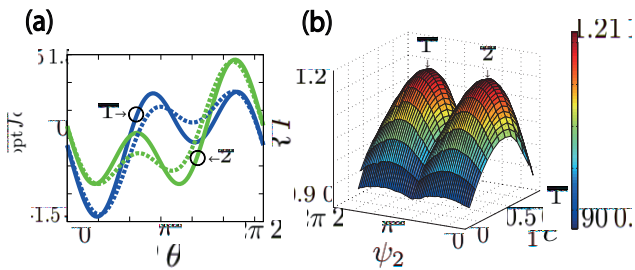}
\caption{\label{cap:GA}Optimal forcing ${\it f}_{{\rm opt}}$ obtained
in the model with the response function of Eq. (\ref{eq10}).
Panel (a) shows all ${\it f}_{{\rm opt}}$, after rescaling for comparison,  obtained by the 
genetic algorithm for $a = 0.95$: green and blue curves for $P=0.5$ with Eq. (\ref{eqnew2}), 
dotted curves for $P=10.0$ with Eq. (\ref{eq1}).
In this genetic algorithm, ${\it f}_{{\rm opt}}$ is searched among all functions of
the form $c_{1}{\rm sin}\theta + c_{2}{\rm sin}(2\theta + \psi_{2})$.
Panel (b) shows the landscape of {\it R} for $a = 0.95$ $( >1/2 )$,
with respect to ($c_{1}$, $\psi_{2}$).
The peaks 1 and 2 of the landscape correspond to the waveforms 1 and 2 
 in panel (a), respectively.}
\end{figure}

\begin{figure}[!htb]
\includegraphics[width=0.9\columnwidth,keepaspectratio]{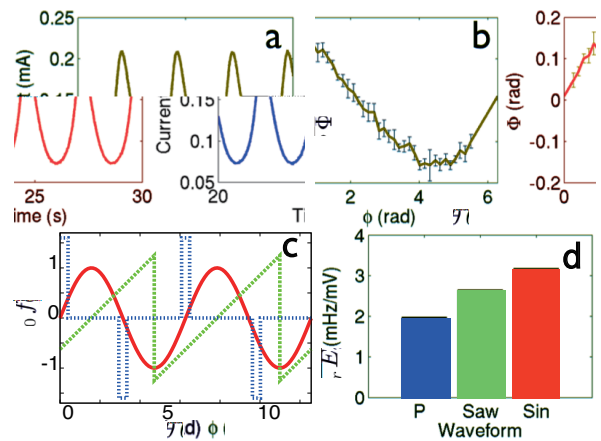}
\caption{\label{cap:exp-hopf}Experiments: optimal entrainment of
smooth oscillators,
$V_{0}=1.100 \hspace{1mm} \textrm{V}$. {(a)} Waveform. {(b)} {Phase response}
curve ({PRC}). The {PRC} was obtained with pulsing
the oscillations with an amplitude of -100 \textrm{mV} and width of 0.1
radian and measuring the phase advance ($\Phi$) in 5 independent experiments.
The error bars indicate standard deviations in the five experiments.
{(c)} Three waveforms, sine-sawtooth-pulse, with which entrainments
are tested. {(d)} Reduced entrainabilities
of the three waveforms: P: pulse, Saw: sawtooth, Sin: sine. The detuning
$\Delta \omega / \omega$ was set to $\pm5$\%; at each
detuning two experiments were carried out. }
\end{figure}

We have also verified these theoretical predictions, by using a
standard genetic algorithm which numerically searches for
 ${\it f}_{{\rm opt}}$.
Figure 1(a) shows all ${\it f}_{{\rm opt}}$ obtained by the algorithm, for 
both cases of 
full (Eq. (\ref{eq1})) and averaged (Eq. (\ref{eqnew2})) phase models, 
with $\omega = 10$, $a = 0.95$ $(>1/2)$ and $P = 0.5$ or $P = 10.0$.
The numerical algorithm found the exact same optimal waveform 
as predicted by the theory, for both cases up to around $P = 2.0$. 
The numerical value of the 
locking range for these 
optimal solutions (maximum in $R$ landscape in Figure 1(b)) is also 
the same as predicted by the theory; the non-generic optimal solution 
performs about $23.1\%$ better than the simple sinusoidal 
(generic optimal) forcing, and also it performs about {$89.8\%$} better than the pulse forcing in Fig. \ref{cap:exp-hopf}(c).
Beyond $P=2.0$, a small discrepancy appears in the optimal waveforms obtained with the 
genetic algorithm.
However, the shape of bimodal landscapes of $R[f]$ in Fig. \ref{cap:GA}(b) is preserved up to around $P = 10.0$ for the case of Eq. (\ref{eq1}), 
which suggests that, at least in this particular example, beyond the strictly weak forcing limit, the theoretical prediction of optimal waveform could be used as an 
initial candidate that can be further optimized with other techniques. \\
\hspace{3mm}The theoretical method for obtaining optimal forcing waveform
was demonstrated in chemical experiments with Ni electrodissolution.
The experiments were carried out in a standard three-electrode electrochemical
apparatus with a 1 mm diameter Ni wire as working, a 1.57 mm diameter
Pt coated Ti rod as counter, and
$\textrm{Hg}/\textrm{H}\textrm{g}_{2}\textrm{S}\textrm{O}_{4}/
\textrm{K}_{2}\textrm{S}\textrm{O}_{4}\textrm{(sat)}$
as reference electrode. The potential of the Ni wire was set to a circuit potential $V=V_{0}+A{f_{0}}({\theta})$,
where $V_{0}$ is a base potential, $A$ is the amplitude of
forcing, and $f_{0}(\theta)$ is the normalized forcing waveform 
with $<f_0^2>=0.5$.
The current, proportional to dissolution rate was measured
by the potentiostat. The frequency of the current oscillations
was determined with the Hilbert transform method \cite{sync-book}. The experiments
were carried at $10\ ^\textrm{o}\textrm{C}$ in 3 mol/L sulfuric acid solution.
Further experimental details are given in previous studies {\cite{Istv, Istv2}}.
When {a resistor of} $1 \hspace{1mm} \textrm{kOhm}$ is attached to the Ni
electrode, oscillations
arise through Hopf bifurcation at $V_{0}\approx1.07 \hspace{1mm} \textrm{V}$ due to the hidden negative differential resistance of the electrochemical
system \cite{Krischer}. %

In accordance with previous experimental results \cite{Istv2}, at $V_{0}=1.100 \hspace{1mm} \textrm{V}$, slightly above the bifurcation point, the oscillations
are smooth with a phase response curve that contains predominantly first harmonic components (see {Figs. \ref{cap:exp-hopf}(a) and \ref{cap:exp-hopf}(b)}). 
The experiments were carried 
out at fixed forcing frequencies and the amplitude of the waveform was increased
until entrainment occurs where the critical amplitude, $A_{c}$, was
recorded. 
To compare the forcing
capability of different waveforms we determined the reduced entrainabilities
as $E_{r}=|\Delta \omega|/(A_{c}\sqrt{<f_0^2>})$. 
The average value of the reduced 
entrainability for fixed positive and negative detunings 
is proportional to the $R$ value; therefore, it is a useful quantity 
to characterize the entraining capability of a waveform. In a typical set of experiments 5 \%
detuning was applied, i.e.,  $\Delta \omega / \omega = \pm 0.05$; it was confirmed that 
smaller detuning (2 \%) gives $E_r$ values within 5 \% relative error; experiments with larger detunings resulted in 
large changes of oscillation waveforms and thus the weak forcing assumption of the
theory was not satisfied. 

We have tested three waveforms with smooth oscillators; sine, sawtooth,
and pulse waves are shown in {Fig. \ref{cap:exp-hopf}(c)}. The optimal
waveform was determined by the above algorithm {\cite{FFT}}{;} only generic waveform
exists and because the {phase response curve} has very small third and higher harmonics
(less than 1\%) we can consider the sine waveform as optimal within experimental
error. The entrainabilities in {Fig. \ref{cap:exp-hopf}(d)} show that,
as predicted, the {sine} waveform indeed outperforms the sawtooth and
pulse waves{.}

\begin{figure}[!htb]
\includegraphics[width=0.85\columnwidth,keepaspectratio]{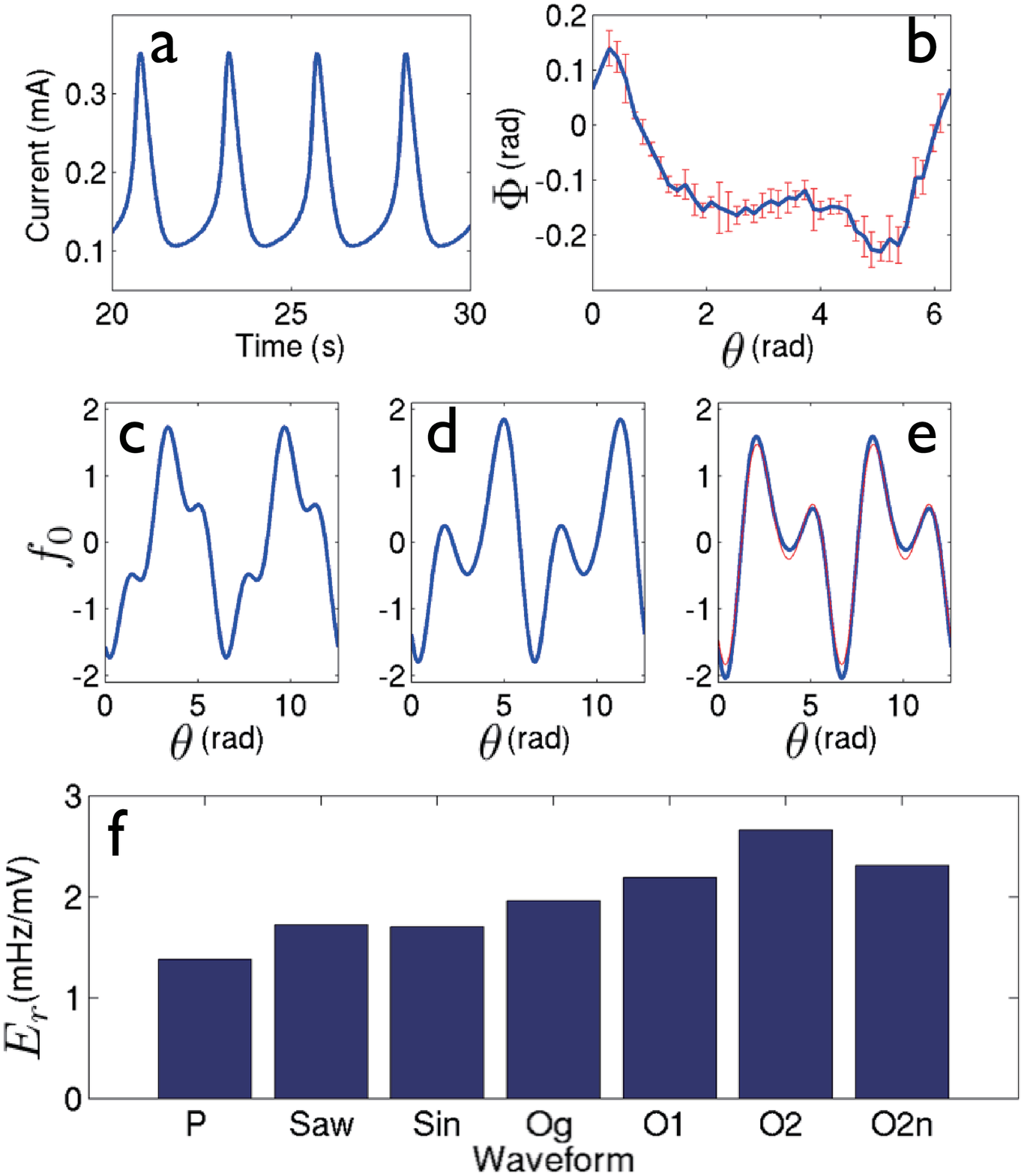}
\caption{\label{cap:exp-1200} Experiments: optimal entrainment of moderately
relaxational oscillators, $V_{0}=1.200 \hspace{1mm} \textrm{V}$. {(a)} Oscillation waveform.
{(b)} {Phase response curve}. The {PRC} was obtained similarly to that
in Fig. \ref{cap:exp-hopf}. {(c--e)} Optimal waveforms. {(c)} Generic optimal waveform (Og). 
{(d)} Optimal waveform 1 (O1). {(e)} Thick solid line: Optimal waveform
2 (O2). Thin line: noisy optimal waveform 2 (O2n). {(f)} Reduced entrainabilities
of the tested waveforms. The $E_{r}$values are averages of experiments
with $\pm2$\% and $\pm5$\% detunings. }
\end{figure}

When $V_{0}$ is increased to 1200 mV, the oscillations become moderately
relaxational; the waveform and the phase response curve exhibits higher harmonics 
(see {Figs. \ref{cap:exp-1200}(a) and \ref{cap:exp-1200}(b)}).
 The analysis has revealed
that three optimal waveforms exist: one generic, and two non-generic
waveforms given in {Figs. \ref{cap:exp-1200}(c--e)}, respectively.
In {Fig. \ref{cap:exp-1200}(e)} we also show a waveform that is obtained
by adding Gaussian random numbers with standard deviation of 0.1 to
the Fourier coefficients of the most optimal waveform; this waveform
is referred to as noisy optimal 2 waveform.
The reduced entrainabilities follow the theoretical predictions; sawtooth,
sine, and pulse waves are inferior to the entraining power of the
optimal waveforms. As predicted, the generic optimal waveform performs
slightly worse than the non-generic optimal waveforms. The best waveform was
optimal 2; note that the shape of this waveform is strongly 
{nontrivial} and {by the addition of small amount of noise to the 
waveform the $E_{r}$ value decreases}.
Overall, optimal waveform obtained from the theory increased entrainabilities by about $50\%$ to $90\%$ relative to those obtained with sine/pulse/sawtooth waves.  

Construction of optimal waveform for entrainment was proposed and
tested here with a single oscillator. The method, however, can be
extended to a group of interacting oscillators where effects related
to the collective phase response function {\cite{kawa}} shall be
considered. The optimal signal can also be applied in closed-loop
feedback systems along with synchronization engineering {\cite{Istv}} for
seeking optimal target dynamics. A limitation of the methodology is
the requirement for weak forcing so that phase models can be applied.
This limitation leads to an extension of the method where the forcing
signal is limited to small values. These extensions, along with other targets that consider 
stability and basin of attraction, will be considered
in a forthcoming publication. {The proposed methodology provides a
framework for efficient design of entrainment applications in 
electrical circuit technology (e.g., for {injection-locked oscillators}) as well as in biological pacemakers.}

\begin{thebibliography}{99}
\bibitem{Reso} In this Letter we focus on the $1:1$ resonant case. However, 
{} the presented theory is also valid for general {\it m} : {\it n} resonant cases.
\bibitem{R} There may be multiple locking ranges for ${\it f}$.
For graphical examples, see \cite{kura}, p. 64. In our setting,
the largest range
should be chosen for maximizing $R$ in these situations.
\bibitem{cons} Here $f$ are continuous and have continuous first derivatives, which belong to a function space with the norm $\| \cdot\|$ defined by $\| f\| = {\rm max}|f(\theta)| + {\rm max}| f'(\theta)|$. For details, see Chap. 1 in \cite{Gelf}.
\bibitem{generic}The existence of the generic solution to balancing condition can be proven by partially integrating $Z'(\theta +\pi) Z(\theta)$ resulting in $\langle Z'(\theta + \pi) Z(\theta) \rangle=0$. Note that this solution has $\mathbb{Z}_2$ symmetry.
\bibitem{FFT} Up to three harmonics are considered in the PRC. Addition 
of higher harmonics resulted in very minor changes of the waveform.
\end{thebibliography}

I.Z.K. acknowledges financial support from Research Corporation 
 Cottrel College Science Award. 
H.T.  thanks Dr. Y. Ando and N. Sagayama for their help with genetic algorithms.

\end{document}